\documentclass[]{spie}  

\usepackage{amsmath,amsfonts,amssymb}
\usepackage{graphicx}
\usepackage[colorlinks=true, allcolors=blue]{hyperref}

\title{Modernising Observatory Control: Integrating Legacy
DRAMA with Next-Generation Instrumentation}

\author[a]{Mrunmayi S. Deshpande}
\author[a]{Nuria, P, F, Lorente}
\affil[a]{Australian Astronomical Optics, Macquarie University, Sydney, Australia}
\author[a]{Tony Farrell}
\author[a]{Satyam Mishra}

\authorinfo{Further author information: (Send correspondence to Mrunmayi Deshpande.)\\: E-mail: mrunmayi.deshpande@mq.edu.au}

\begin{document} 
\maketitle

\begin{abstract}
The DRAMA framework was originally developed at the Anglo-Australian Observatory (AAO) in the early 1990s, with its first release in 1992. Designed in C and C++, DRAMA provides a distributed, message-based control architecture that has powered major instruments such as the 2dF fibre positioner, fibre-fed spectrographs like AAOmega and HERMES, and the IRIS2 infrared imager on the Anglo-Australian Telescope, and the TAIPAN instrument on the UK Schmidt Telescope. Despite its age, DRAMA remains a robust and well-structured system whose real-time behaviour and modular task oriented design continue to be valuable for astronomical instrumentation.
As new instruments are being developed at the AAO, we are extending and adapting the DRAMA environment to meet the current software engineering and deployment requirements. Our recent improvements include containerising DRAMA using Docker and a modern Python interface that enables integration without modifying the underlying codebase. This approach preserves the reliability of the original framework, and allows new instruments to adopt DRAMA within more flexible and maintainable ecosystems. At present, DRAMA continues to play an active role in our development efforts: it is being used for the newly built KSPEC spectrograph on the KMTNet telescope, and also forms the control architecture for the high-resolution spectrograph being developed for the Devasthal Optical Telescope.

\end{abstract}


\section{INTRODUCTION}
\label{sec:intro}
The DRAMA (Distributed Real-Time AAO Monitor for Astronomy) API was designed to meet the AAO's requirements for a fast, distributed environment for writing Instrumentation Control Systems~\cite{farrell1993drama}. Originally implemented in 1992 by the then Anglo-Australian Observatory (AAO), the framework was heavily influenced by the tasking structures of the earlier Starlink ADAM environment~\cite{allan1992adam}. It was engineered to manage large, heterogeneous networks of computers spanning Unix, macOS, and hard real-time VxWorks kernels. Instruments demanded highly predictable real-time performance in machines with limited resources. Over more than three decades, it has successfully powered mission-critical systems, including the 2dF fibre positioner~\cite{lewis2002twodf}, the AAOmega and HERMES spectrographs, and the IRIS2 infrared imager on the 3.9m Anglo-Australian Telescope (AAT). It was also deployed on the UKIRT, JCMT, WHT, Mayall and Blanco telescopes, and various others.  It was designed specifically for writing instrument control software but has also been deployed for the data reduction systems such as 2dFdr~\cite{2dfdr}. The framework is highly optimised to meet the strict performance demands of astronomical instrumentation, focusing heavily on fast, deterministic inter-task messaging and distributed system. To achieve these high-performance requirements, the framework provides three core operational capabilities:

\begin{itemize}
  \item Standardised Task Structure: A standard task  structure is provided by the Generic Instrumentation Task (GIT) on top of a very flexible layer allowing others to implement their own architecture.
  \item Network Transparency: DRAMA abstracts the underlying network layer, allowing tasks running on separate physical machines (e.g., control computers, data acquisition, or telescope front-ends) to communicate seamlessly as if they were local.
  \item High Performance: Minimises latency in command processing and data transport, which is critical for instrument control.
\end{itemize}

DRAMA coordinates distributed communication using independent executable programs designated as Tasks. These Tasks interact using an external messaging model consisting of Actions (commands invoked via OBEY messages) and Parameters (state variables). Early implementations predated standard ANSI C and uniform operating system threading models. As a result, tasks historically relied on single-threaded, cooperative multi-tasking. Under this model, action handlers were strictly required to execute rapidly and return control to the main message-reading loop. Although structurally efficient, this cooperative staging forced developers to manage complex, non-linear state machines. Although highly useful,this suffered drawbacks, including lack of modern exceptional framework.  n frameworks, which prompted a major architectural shift with the release of DRAMA2~\cite{farrell2017drama2}. Leveraging the standardised threading, synchronisation and memory management introduced in C++11~\cite{iso2011cpp11} and C++14~\cite{iso2014cpp14}, DRAMA2 re-engineered the environment into a modern object-oriented system. By introducing native execution threads per action managed by abstract handler classes, DRAMA2 eliminates the need for developers to manually manage intricate state machines. This allowed them to write more natural, linear sequence code without blocking the underlying network message loop.

In the era of next-generation instrumentation, the software engineering demands of instrument and telescope control systems have shifted from basic execution models to deployment flexibility, long-term cross-platform compliance, and high-level scripting integration. This paper details our recent engineering efforts to build directly upon the foundational DRAMA2 architecture. We present a modernised approach that introduces containerisation, robust automated continuous integration (CI/CD), strict compiler standards compliance, and native high-level language bindings.

\section{ARCHITECTURAL OVERVIEW}

The architecture and legacy foundations of DRAMA is highly modular, splitting its responsibilities into distinct layers. At its core, DRAMA is based on a fundamental architectural rule: Tasks do not communicate directly with each other via operating system primitives. Instead, all interactions are deconstructed and executed by the core DRAMA libraries.

\begin{figure} [ht]
\begin{center}
\begin{tabular}{c} 
\includegraphics[height=10cm]{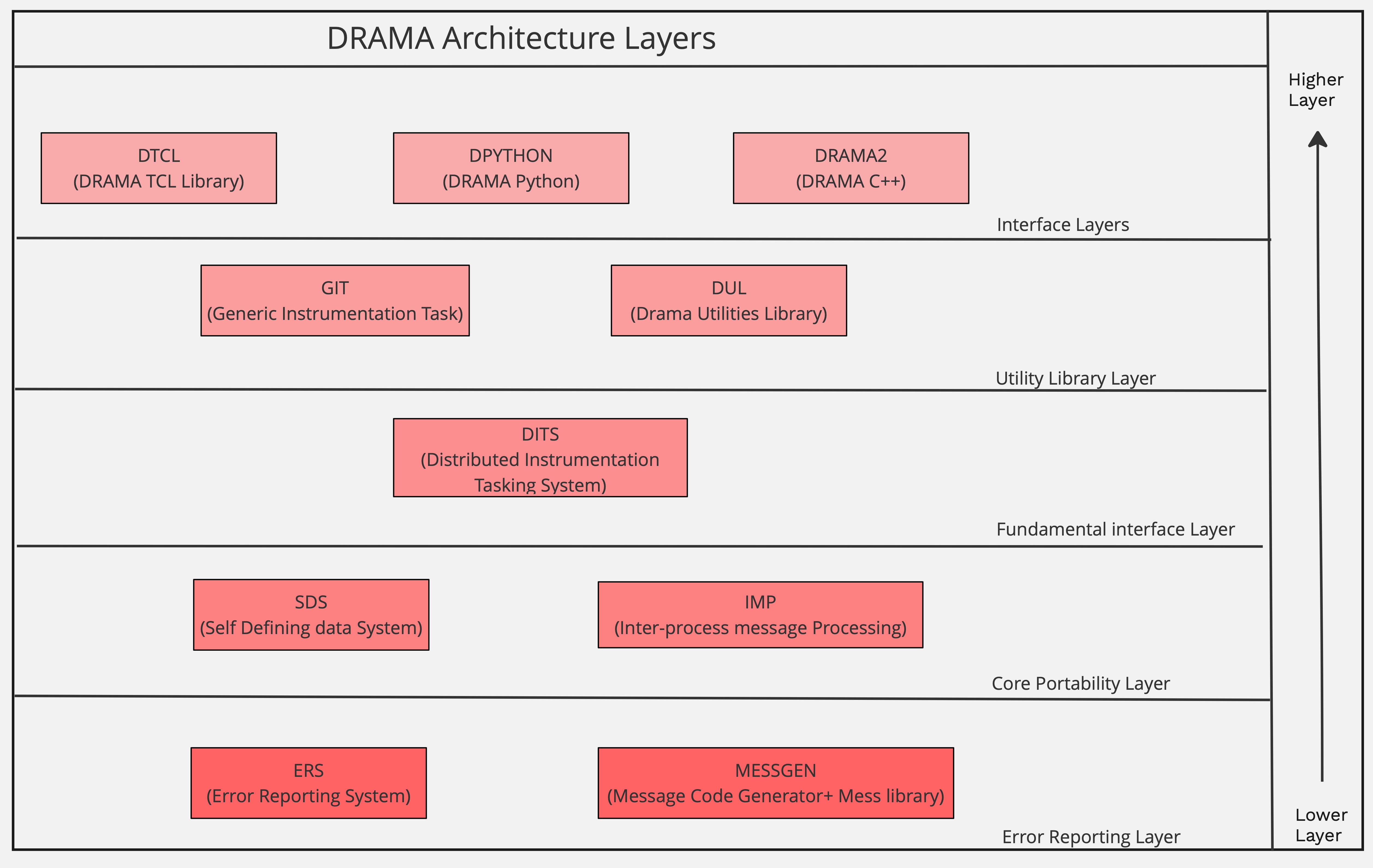}
\end{tabular}
\end{center}
\caption[example] 
{ \label{fig:drama_img} 
Drama Architecture showing the five functional layers and their relationships}
\end{figure} 
\subsection{The Five Functional Layers}
\label{subsec:layers}

\begin{description}
    \item[1. Error Reporting Layer (MESSGEN \& ERS)] \hfill \\
    The backbone of the DRAMA environment is its error handling pipeline, which operates via two core components:
    \begin{description}
        \item[\texttt{MESSGEN} (Message Generator):] A specialised utility compiler and associated library that translates human-readable error definition files into standard C/C++ header files and source code. It assigns unique, structured integer status codes to specific subsystems, ensuring that error codes never clash across different instrumentation packages.
        \item[\texttt{ERS} (Error Reporting System):] This is a contextual error logging infrastructure. Unlike simple logging routines, \texttt{ERS} allows lower level code to report basic error information while the upper level code can add context information or choose not to report the error.
    \end{description}

    \item[2. Core Portability Layer (SDS \& IMP)] \hfill \\
    This layer abstracts operating system dependencies, memory access patterns, and network topology to ensure seamless cross-platform deployment:
    \begin{description}
        \item[\texttt{SDS} (Self-Defining Data System):] A highly optimised data structure compiler and runtime library. \texttt{SDS} allows complex data types (such as multi-dimensional arrays, instrument configurations, or exposure metadata) to be packed into binary format that minimises translation when moving data between machines of different architecture (e.g. byte order). 
        \item[\texttt{IMP} (Interprocess Message Passing):] The underlying point-to-point message protocol. The \texttt{IMP} sub-system can move data between tasks on the same machine or on different machines with minimal, and in some cases, no copying of data, making it suitable for transferring large image files.
    \end{description}

    \item[3. Fundamental Interface Layer (DITS)] \hfill \\
    The Distributed Instrumentation Task System (\texttt{DITS}) represents the core behavioural specification of a DRAMA. It translates \texttt{IMP} messages into formal DRAMA semantics:
    \begin{itemize}
        \item \texttt{DITS} defines how a program registers itself as a uniquely named \textbf{Task} on the network.
        \item It implements the foundational execution loop that intercepts incoming message traffic and dispatches control to specific user-defined \textbf{Actions}, manages active states, and processes asynchronous interrupts. All legacy and modern DRAMA subsystems ultimately route through the \texttt{DITS} processing loop.
        \item Provides a wide range of features for supporting Instrument Tasks in particular. E.g., support for asynchronous input events, creating GUIs, aborting operations, and logging.
    \end{itemize}

    \item[4. Utility Library Layer (GIT \& DUL)] \hfill \\
    To avoid rewriting boilerplate code for every new instrument, this layer provides architectural standardisation and common software patterns:
    \begin{description}
        \item[\texttt{GIT} (Generic Instrumentation Task):] A support for the
AAO’s generic instrumentation software. This facilitates the building tasks with consistent
interfaces in an object-oriented way. \texttt{GIT} defines a mandatory, universal subset of actions (such as \texttt{INITIALISE}, \texttt{RESET}, \texttt{POLL}, and \texttt{EXIT}) and standard parameters (like \texttt{simulation\_mode}) that \textit{every} instrument task must support, ensuring that a centralised observatory control system can interact with any device. External users of DRAMA have used similar approaches to implement their own local standards.
        \item[\texttt{DUL} (DRAMA Utilities Library):] A collection of high-level utility routines that simplify recurring development challenges. \texttt{DUL} provides wrappers around various common operations in the lower levels. Operations such as sending messages to other tasks are presented in a simpler interfaces, but maybe not with the efficiency and/or flexibility provided by the underlying DITS routines.
    \end{description}

    \item[5. Interface Layers (DTCL, DPYTHON \& DRAMA2)] \hfill \\
    The uppermost layer provides language-specific bindings, enabling developers to build applications using their high level language of choice while maintaining native access to the high-performance C-layer backbone. Complete technical documentation and the core distribution files for these subsystems are maintained through the official DRAMA portal~\cite{drama_website}.
    \begin{description}
        \item[\texttt{DTCL} (DRAMA Tcl/Tk):] The legacy scripting interface, historically used to bind DRAMA tasks to graphical user interfaces and operational command scripts via the Tcl language.
        \item[\texttt{DPYTHON} (DRAMA Python):] The modern high-level scripting wrapper. It maps DRAMA actions and \texttt{SDS} structures into native Python objects, enabling rapid user-interface scripting, automated test-suite orchestration, and direct data integration with modern scientific software libraries (numpy, scipy) enabling rapid user-interface scripting.
        \item[\texttt{DRAMA2}:] The modern C++ execution environment. Leveraging the standardised multi-threading capabilities, synchronisation primitives, and memory safety semantics of C++11 and C++14. By providing native execution threads per action (\texttt{TAction}) it allows engineers to write more natural, linear sequence code that interacts with hardware, without blocking or stalling the underlying network message loop.
    \end{description}
\end{description}

\section{MODERNISATION AND DEV-OPS INTEGRATION}
\label{sec:devops_integration}

To lower the barrier to entry for new developers and simplify deployment across diverse computing environments, a concentrated effort was made to integrate modern DevOps workflows and containerised infrastructure into the DRAMA ecosystem. The framework continues to control frontline astronomical instrumentation, including the Veloce ultra-precise radial velocity spectrograph~\cite{Veloce} and the Hector multi-object integral field spectrograph~\cite{Hector} at the Anglo-Australian Telescope (AAT), as well as the Taipan instrument~\cite{TAIPAN} on the UK Schmidt Telescope. It is now being actively deployed for next-generation systems, as detailed in Section~\ref{sec:next-gen}. 

While the legacy software originally managed data from $1024 \times 1024$ pixel detectors, the updated system seamlessly handles modern $10,000 \times 10,000$ pixel detector formats with zero modifications to the core application logic. Although upgraded physical computing infrastructure accounts for some performance gains, the DRAMA core architecture demonstrates its inherent efficiency by maximising data transfer between independent programs. Following the institutional transition of Australian Astronomical Optics (AAO) to Macquarie University, the framework is now used in active development by both the AAO and ANU. This has significantly expanded core developer base as compared to the previous decade.

\subsection{Containerisation and Docker-Based Environments}
Deploying long-lifecycle instrumentation software in modern operating systems frequently introduces brittle dependency chains, mismatches in the operating system, and compilation vulnerabilities. We resolved these deployment hurdles by containerising the complete DRAMA environment using Docker. This standardised base DRAMA Docker image is published to a centralised container registry, allowing it to be dynamically pulled by any instrument control repository. This architecture provides a deterministic, isolated runtime environment for instrument control applications during local development, automated testing, and laboratory hardware simulation loops. This containerised layer acts as an architectural abstraction layer on top of the DRAMA2 framework, establishing robust cross-platform reproducibility.

\subsection{Automated CI/CD Pipelines and Build Efficiency}
Migrating the codebase to a centralised GitLab repository enabled the execution of contemporary continuous integration and continuous deployment (CI/CD) pipelines. Built entirely around the GitLab CI/CD runner engine, the compilation ecosystem utilises the standardised DRAMA Docker base image to guarantee deterministic, reproducible binary builds across all target environments. 

The historical configuration manager (\texttt{dmkmf}) has been refactored to dynamically verify localised environment flags and system paths. Automated pipelines continuously compile the core system utilities and dynamically package the automated Python interface (\texttt{dpython}) bindings upon every commit to the repository. All recently developed AAO instrumentation software will then be rebuilt against the new commit and likewise packaged up. Automated regression testing is being added to all new and modified systems as part of development processes. 

\subsection{Documentation and Release Automation}
While DRAMA has always automated much of its documentation generation, the deployment of updates was time consuming. To support a collaborative open-source model and streamline cross-institutional development, the technical documentation pipeline is fully automated. Legacy documentation files and code comments are programmatically parsed and published automatically to a secure web front-end via GitLab Pages. This automated deployment model ensures that any API modifications, exception-handling frameworks, or structural architectural updates are immediately compiled and searchable by engineers and instrument scientists globally.

\section{NEXT-GENERATION INSTRUMENTATION CASE STUDIES}
\label{sec:next-gen}
\subsection{The KSPEC Spectrograph}
The KSPEC Spectrograph is scheduled for commissioning at the Siding Spring Observatory in Australia as part of a collaborative effort led by a consortium of Korean institutes, including Seoul National University (SNU), the Korea Astronomy and Space Science Institute (KASI), and the Korea Institute for Advanced Study (KIAS) on the KMTNet telescope array. Designed and built by Australian Astronomical Optics (AAO), KSPEC represents optimised evolution of the TAIPAN spectrograph architecture originally deployed on the UK Schmidt Telescope (UKST)~\cite{mcgregor2026kspec}. 

The software execution relies comprehensively on the modernised DRAMA ecosystem to orchestrate these capabilities. The architecture leverages the DRAMA infrastructure to handle the underlying low-level hardware communication in real time, allowing high-level software controls. It coordinates intricate spectrograph operations including taking observation runs, data recording, detector control, and spectrograph movements. At the centre of this runtime environment is a specialised control task that drives the underlying operations of the spectrograph mechanism, the CCD detector, and the data-ingestion pipelines using native DRAMA actions. Following the successful completion of hardware fabrication, sub-assembly, and alignment phases, the instrument is actively tracking toward first light by the end of 2026.

\subsection{The Devasthal Optical Telescope (DOT) High-Resolution Spectrograph}
The Devasthal Optical Telescope High Resolution Spectrograph (DOT-HRS) is a next-generation instrument designed for installation on  the 3.6-m (DOT) at the Aryabhatta Research Institute of Observational Sciences (ARIES)~\cite{pandey2019dothrs}. Built to support a broad spectrum of research applications, the bench-mounted, fiber-fed Echelle spectrograph operates in both High Resolution (R~$\approx$~80,000) and High Efficiency (R~$\approx$~40,000) modes across a wide spectral baseline from 380 to 850~nm.
Currently in development, DOTHRS uses DRAMA as its core architecture and seamlessly integrates with OPC UA, QT, and Python interfaces. In the DOTHRS project, DRAMA middleware is used for inter-module communication, error detection and propagation, and logging across the instrument software architecture. DRAMA integrates along with the OPC UA framework~\cite{opcua_standard} to handle large image transfers and to implement complex control algorithms in a distributed control environment.
DRAMA components are used to create FITS files, including metadata for header information from various subsystems. Other DRAMA components that match the DOT/HRS requirements have been reused or forked for the HRS software.
The Control Task uses DRAMA to communicate with the technical cameras and the CCD controller. DRAMA software is also used for camera control and for recording scientific data generated by the instrument.

\section{CONCLUSIONS AND FUTURE WORK}
The evolution of the DRAMA framework from a 1992 single-threaded cooperative multi-tasking environment to the modern, containerised, multi-threaded DRAMA2 ecosystem demonstrates its foundational architectural resilience. By successfully bridging legacy stability with modern software engineering paradigms, the framework proves that older, highly specialised instrumentation control software does not need to be abandoned to satisfy modern operational requirements. 

Through the integration of a unified GitLab CI/CD pipeline, native Docker based environments, and complete automated API\cite{aao_drama_repo} documentation through GitLab Pages\cite{drama_gitpages} , the barrier to entry for the next generation of engineers and instrument scientists has been drastically reduced. Ultimately, the real-world viability of this modernised architecture is firmly demonstrated by its active deployment at major global facilities: successfully driving real-time automated telescope and instrument operations, accelerating data pipelines, and securing long-term software sustainability for future instrumentation frontiers. DRAMA has been well-designed from the beginning, enhanced with every instrument and observatory deployment is continuously developing and being used for next-generation astronomy instruments effectively

\appendix    

\acknowledgments 
The authors gratefully acknowledge the contributions of the many developers and engineers who have worked on DRAMA over the years, as well as the teams responsible for the astronomical instruments and telescopes that have adopted and advanced its use. Their efforts have been instrumental in the development, testing, commissioning, and continued success of the DRAMA software framework.

\bibliography{report} 

@string{spie = {Proc. SPIE}}

@inproceedings{farrell1993drama,
  author    = {Farrell, T. J. and Bailey, J. A. and Shortridge, K.},
  title     = {{DRAMA: An Object-Oriented Distributed Instrumentation Control System}},
  booktitle = {Bulletin of the American Astronomical Society},
  volume    = {25},
  pages     = {915},
  year      = {1993}
}

@inproceedings{farrell2017drama2,
  author    = {Farrell, Tony and Shortridge, Keith},
  title     = {{DRAMA2 - DRAMA for the Modern Era}},
  booktitle = {Astronomical Data Analysis Software and Systems XXVI},
  editor    = {Molano, F. and Guzman, J. C.},
  series    = {ASP Conference Series},
  volume    = {512},
  pages     = {123},
  year      = {2017},
  publisher = {Astronomical Society of the Pacific}
}

@article{lewis2002twodf,
  author    = {Lewis, I. J. and Cannon, R. D. and Taylor, K. and Glazebrook, K. and Bailey, J. A. and Baldry, I. K. and Bridges, T. J. and Bridges, G. and Bridges, D. J. and Cram, L. and d'Amico, F. and Fine, S. and Frost, G. and Glendinning, R. and Gray, P. M. and Gillingham, P. R. and Lankshear, A. and Lucey, J. R. and Maugham, B. and Parker, Q. A. and Saunders, W. and Smith, S. and Smith, G. and Willis, M.},
  title     = {{The 2dF Galaxy Redshift Survey: Design and operation of the 2dF facility}},
  journal   = {Monthly Notices of the Royal Astronomical Society (MNRAS)},
  volume    = {333},
  number    = {2},
  pages     = {279--299},
  year      = {2002},
  doi       = {10.1046/j.1365-8711.2002.05333.x}
}

@INPROCEEDINGS{2dfdr,
       author = {{Farrell}, Tony and {Birchall}, Michael and {Croom}, Scott and {Lidman}, Chris},
        title = "{2dFdr - One million Spectra and Counting}",
    booktitle = {Astronomical Data Analysis Software and Systems XXVI},
         year = 2019,
       editor = {{Molinaro}, Marco and {Shortridge}, Keith and {Pasian}, Fabio},
       series = {Astronomical Society of the Pacific Conference Series},
       volume = {521},
        month = oct,
        pages = {675},
       adsurl = {https://ui.adsabs.harvard.edu/abs/2019ASPC..521..675F}
}

@inproceedings{allan1992adam,
  author    = {Allan, P. M.},
  title     = {{The ADAM Software Environment}},
  booktitle = {Astronomical Data Analysis Software and Systems I},
  editor    = {Worrall, D. M. and Biemesderfer, C. and Barnes, J.},
  series    = {ASP Conference Series},
  volume    = {25},
  pages     = {126},
  year      = {1992},
  publisher = {Astronomical Society of the Pacific}
}

@techreport{iso2011cpp11,
  author      = {{ISO/IEC JTC 1/SC 22/WG 21}},
  title       = {{ISO/IEC 14882:2011: Information technology --- Programming languages --- C++}},
  institution = {International Organization for Standardization},
  type        = {Standard},
  number      = {ISO/IEC 14882:2011},
  year        = {2011},
  url         = {https://www.iso.org/standard/50372.html}
}

@techreport{iso2014cpp14,
  author      = {{ISO/IEC JTC 1/SC 22/WG 21}},
  title       = {{ISO/IEC 14882:2014: Information technology --- Programming languages --- C++}},
  institution = {International Organization for Standardization},
  type        = {Standard},
  number      = {ISO/IEC 14882:2014},
  year        = {2014},
  url         = {https://www.iso.org/standard/64029.html}
}

@INPROCEEDINGS{Veloce,
       author = {{Gilbert}, James and {Bergmann}, Christoph and {Bloxham}, Gabe and {Boz}, Robert and {Brookfield}, Robert and {Carkic}, Tom and {Carter}, Brad and {Case}, Scott and {Churilov}, Vladimir and {Ellis}, Michael and {Gausachs}, Gaston and {Gers}, Luke and {Gray}, Doug and {Herrald}, Nicholas and {Ireland}, Michael and {Jones}, Damien and {Kripak}, Yevgen and {Lawrence}, Jon and {O'Brien}, Ellie and {Price}, Ian and {Robertson}, Matthew and {Schwab}, Christian and {Tinney}, Chris and {Vaccarella}, Annino and {Vest}, Colin and {Wright}, Duncan and {Zhelem}, Ross},
        title = "{Veloce Rosso: Australia's new precision radial velocity spectrograph}",
    booktitle = {Ground-based and Airborne Instrumentation for Astronomy VII},
         year = 2018,
       editor = {{Evans}, Christopher J. and {Simard}, Luc and {Takami}, Hideki},
       series = {Society of Photo-Optical Instrumentation Engineers (SPIE) Conference Series},
       volume = {10702},
        month = jul,
          eid = {107020Y},
        pages = {107020Y},
          doi = {10.1117/12.2312399},
archivePrefix = {arXiv},
       eprint = {1807.01938},
 primaryClass = {astro-ph.IM},
       adsurl = {https://ui.adsabs.harvard.edu/abs/2018SPIE10702E..0YG}
}

@INPROCEEDINGS{Hector,
       author = {{Bryant}, Julia J. and {Bland-Hawthorn}, Joss and {Lawrence}, Jon and {Norris}, Barnaby and {Min}, Seong-Sik and {Brown}, Rebecca and {Wang}, Adeline and {Bhatia}, Gurashish Singh and {Saunders}, Will and {Content}, Robert and {Zhelem}, Ross and {Venkatesan}, Sudharshan and {Mohanan}, Mahesh and {Gillingham}, Peter and {Patterson}, Robert and {Robertson}, David and {Pai}, Naveen and {McGregor}, Helen and {Zheng}, Jessica and {Vaughan}, Sam and {Foster}, Caroline and {Leon-Saval}, Sergio and {Croom}, Scott},
        title = "{Hector: a new multi-object integral field spectrograph instrument for the Anglo-Australian Telescope}",
    booktitle = {Ground-based and Airborne Instrumentation for Astronomy VIII},
         year = 2020,
       editor = {{Evans}, Christopher J. and {Bryant}, Julia J. and {Motohara}, Kentaro},
       series = {Society of Photo-Optical Instrumentation Engineers (SPIE) Conference Series},
       volume = {11447},
        month = dec,
          eid = {1144715},
        pages = {1144715},
          doi = {10.1117/12.2560309},
       adsurl = {https://ui.adsabs.harvard.edu/abs/2020SPIE11447E..15B}
}

@inproceedings{TAIPAN,
  author    = {Nicholas F. Staszak and Jon Lawrence and Ross Zhelem and Robert Content and Vladimir Churilov and Scott Case and Rebecca Brown and Andrew M. Hopkins and Kyler Kuehn and Naveen Pai and Urs Klauser and Vijay Nichani and Lew Waller},
  title     = {TAIPAN fibre feed and spectrograph: engineering overview},
  booktitle = {Ground-based and Airborne Instrumentation for Astronomy VI},
  editor    = {Helen J. Hall and Roberto Gilmozzi and Howard K. Marshall},
  series    = {Proceedings of SPIE},
  volume    = {9912},
  pages     = {99125A},
  year      = {2016},
  publisher = {SPIE},
  doi       = {10.1117/12.2233796},
  url       = {https://doi.org/10.1117/12.2233796}
}

@inproceedings{mcgregor2026kspec,
  author    = {McGregor, Helen M. and Hewlett, Max and Cerneaz, Nick and Non, Sambath and Zhelem, Ross and Guzman, Dani and Lou, Summer and Lawrence, Jonathon S. and Waller, Lewis G. and Kim, Jae-Woo and Hwang, Ho Seong},
  title     = {{KSPEC: spectrograph optimization and delivery to the KMTNet telescope}},
  booktitle = {Ground-based and Airborne Instrumentation for Astronomy X},
  series    = {Proc. SPIE},
  volume    = {14149},
  pages     = {14149--393},
  year      = {2026},
  month     = {July}
}

@article{pandey2019dothrs,
  author    = {Pandey, J. C. and Pant, J. and Joshi, Y. C.},
  title     = {{A high-resolution spectrograph for the 3.6-m Devasthal Optical Telescope of ARIES}},
  journal   = {Bulletin de la Soci{\'e}t{\'e} Royale des Sciences de Li{\`e}ge},
  volume    = {88},
  pages     = {43--54},
  year      = {2019},
  doi       = {10.25518/0037-9565.8451}
}

@misc{drama_website,
  author       = {{Australian Astronomical Optics}},
  title        = {{DRAMA: Distributed Real-Time Monitor for Astronomy}},
  howpublished = {\url{https://drama.aao.org.au}},
  year         = {2026},
  note         = {Accessed: 2026-06-26}
}

@misc{aao_drama_repo,
  author       = {{Australian Astronomical Optics}},
  title        = {{DRAMA: Distributed Data Acquisition Environment Core Repository}},
  howpublished = {\url{https://dev.aao.org.au/rds/drama/drama}},
  year         = {2026},
  note         = {Accessed: \today},
  organization = {Macquarie University}
}

@techreport{opcua_standard,
  author      = {{OPC Foundation}},
  title       = {{OPC Unified Architecture --- Part 1: Overview and Concepts}},
  institution = {OPC Foundation},
  type        = {Specification},
  number      = {Release 1.05},
  year        = {2022},
  url         = {https://opcfoundation.org/developer-tools/specifications-unified-architecture}
}

@misc{drama_gitpages,
  author       = {{Australian Astronomical Optics}},
  title        = {{DRAMA: Distributed Real-Time Monitor for Astronomy}},
  howpublished = {\url{https://rds.survey.org.au/drama/drama/}},
  year         = {2026},
  note         = {Accessed: 2026-06-26}
}
\bibliographystyle{spiebib} 

\end{document}